\documentclass[journal]{IEEEtran}
\usepackage{cite}
\usepackage{amsmath,amssymb,amsfonts}
\usepackage{algorithmic}
\usepackage{graphicx}
\usepackage{textcomp}
\usepackage{xcolor}
\usepackage{multicol}
\usepackage{balance}
\usepackage{hyperref}
\usepackage{soul}
\usepackage{longtable}

\usepackage{array}
\usepackage{ragged2e}
\usepackage{makecell}
\usepackage{amsthm}
\usepackage{amsfonts}
\usepackage{color}
\usepackage{soul}
\usepackage{tabularx}
\usepackage{adjustbox}
\usepackage{adjustbox,lipsum}
\usepackage{rotating}
\usepackage{xcolor,colortbl}
\usepackage{subfiles}
\usepackage{multicol}
\usepackage{stfloats}
\usepackage{url}
\usepackage{xurl}
\definecolor{Gray}{gray}{0.9}
\usepackage{tikz}

\begin{document}

\title{Blockchain for the Metaverse: A Review}

\author{Thippa~Reddy~Gadekallu,~\IEEEmembership{Senior Member,~IEEE,} Thien~Huynh-The,~\IEEEmembership{Member,~IEEE,} Weizheng~Wang, Gokul~Yenduri, Pasika Ranaweera,~\IEEEmembership{Member,~IEEE,} 
Quoc-Viet~Pham,~\IEEEmembership{Member,~IEEE,} Daniel~Benevides~da~Costa,~\IEEEmembership{Senior~Member,~IEEE,} and Madhusanka~Liyanage,~\IEEEmembership{Senior~Member,~IEEE}
\thanks{Thien Huynh-The is with the ICT Convergence Research Center, Kumoh National Institute of Technology, Gumi, Gyeongsangbuk-do 39177, Korea (e-mail: thienht@kumoh.ac.kr).}
\thanks{Thippa Reddy Gadekallu and Gokul Yenduri are with the School of Information Technology, Vellore Institute of Technology, Tamil Nadu- 632014, India (e-mail: \{thippareddy.g, gokul.yenduri\}@vit.ac.in).}

\thanks{Weizheng Wang is with the Department of Computer Science, City University of Hong Kong, Hong Kong (e-mail: weizheng.wang@ieee.org).}
\thanks{Pasika Ranaweera is with the School of Computer Science, University College Dublin, Ireland (e-mail: pasika.ranaweera@ucdconnect.ie).}
\thanks{Quoc-Viet Pham is with the Korean Southeast Center for the 4th Industrial Revolution Leader Education, Pusan National University, Busan 46241, Korea (e-mail: vietpq@pusan.ac.kr).}
\thanks{Daniel Benevides da Costa is with the AI $\&$ Telecom Research Center, Technology Innovation Institute, Abu Dhabi, UAE, and Future Technology Research Center, National Yunlin University of Science and Technology, Taiwan, R.O.C. (e-mail: danielbcosta@ieee.org).}
\thanks{Madhusanka Liyanage is with the School of Computer Science, University Collage Dublin, Ireland and Centre for Wireless Communications, University of Oulu, Finland (e-mail: madhusanka@ucd.ie and madhusanka.liyanage@oulu.fi).}
}

\maketitle

\begin{abstract}
Since Facebook officially changed its name to Metaverse in Oct. 2021, the metaverse has become a new norm of social networks and three-dimensional (3D) virtual worlds. The metaverse aims to bring 3D immersive and personalized experiences to users by leveraging many pertinent technologies.
Despite great attention and benefits, a natural question in the metaverse is how to secure its users' digital content and data. In this regard, blockchain is a promising solution owing to its distinct features of decentralization, immutability, and transparency. To better understand the role of blockchain in the metaverse, we aim to provide an extensive survey on the applications of blockchain for the metaverse. We first present a preliminary to blockchain and the metaverse and highlight the motivations behind the use of blockchain for the metaverse. Next, we extensively discuss blockchain-based methods for the metaverse from technical perspectives, such as data acquisition, data storage, data sharing, data interoperability, and data privacy preservation. For each perspective, we first discuss the technical challenges of the metaverse and then highlight how blockchain can help. Moreover, we investigate the impact of blockchain on key-enabling technologies in the metaverse, including Internet-of-Things, digital twins, multi-sensory and immersive applications, artificial intelligence, and big data. We also present some major projects to showcase the role of blockchain in metaverse applications and services. Finally, we present some promising directions to drive further research innovations and developments towards the use of blockchain in the metaverse in the future.

    
    
    
    
\end{abstract}


\section{Introduction}
\label{Sec:Introduction}


The metaverse is the next phase of digital evolution that can revolutionize the digital adoption to a staggering level and extends the domain of services beyond the standard systems with online access. Digitization of services has become the trend for improving the efficiency in the fields of business, entertainment, education, or any other system that can be integrated with online access over the past few decades. These services and systems were improved to its maximum potential with the capabilities provided with digital systems and online storage/ processing facilities at remote data centres and cloud platforms. With the efficiency, performance, and quality of the service access reaching to its highest potential, the perspective has been shifted towards the consumer experience. Thus, the demand for improved service experience with more interactive capability is ever increasing and service providers are keen on elevating their existing standards to the next level. In fact, consumers are demanding haptic and immersive capabilities with their digital interfacing, where such traits are only possible with the emerging technologies of Virtual Reality (VR), Augmented Reality (AR), Mixed Reality (MR), and Extended Reality (XR) \cite{lee2020unified}. The metaverse is the solution that amalgamate all these pertinent technologies in the global context. This concept creates a simulated digitized environment that can be endured as an immersive virtual world for its prosumers. Users can interact with this virtual eco system through their digital avatars in compliance to the duality principle \cite{lee2021all}. Concretely, the avatars are the virtual embodiments of the users, and has the same legal authority in the metaverse as one's legal rights in the real world; this makes the avatar warranted for any transactions made within the virtual domain and restricts from repudiating any committed action.  The access can be gained by any person having a VR/ AR enabled immersive device, such as a headset or a glass under the minimal capability \cite{bolger2021finding}. On the contrary, full-body haptic bodysuits such as Teslasuit or Holosuit carries the potential to embrace the immersive experience to its peak with the capability to track the motions, extract haptic feedback along with transcended biometrics.

Despite the metaverse being developed and intended to expand the scope of capabilities in social media, its potential for other industrial, commercial, societal, educational, medical, military, and governmental sectors are immense. Lack of immersive  experience is a well-known drawback with the online remote access and control systems. Specially in the instances of controlling Supervisory Control and Data Acquisition (SCADA) or Programmable Logic Controller (PLC) based remote automation systems \cite{thepmanee2022implementation}, fitting on apparels, perception in commercial real-estate or architecture, understanding Three-Dimensional (3D) visualization in medical/ engineering/ or architectural education, remote controlling of unmanned aerial/ naval/ or ground vessels, experiencing digital entertainment beyond the 2 dimensions are areas that required more innovation. Though AR and VR technologies offered standalone solutions for these areas, an all-in-one platform or an environment was lacking to combine these tools. The metaverse offers this digital eco-system to the world, and widely open the scope of possibilities beyond measure. The concept of Digital Twins (DTs) empowers the remote operation and controlling of machines or vehicles with improved visualization and coordination, and benefit both industrial and military sectors \cite{ramu2022federated}. Three-dimensional visualization is leading for better accuracy and understanding of the context that benefit both educational and entertainment applications. In addition, novel directives such as AR based remote robotic controlling, AR based remote surgery are achievable with the metaverse platform \cite{ranaweera2020novel}. Further, the concept such as cryptocurrency \cite{bouri2021quantile}, digital-biometrics \cite{bisogni2021ecb2}, and explainable artificial intelligence (XAI) \cite{wang2021explainable} are facing unavoidable challenges when implementing them in the real-world; with the issues of integrating to existing systems, compatibility, inter-operability, legal, and ethical discrepancies. As the metaverse is a newly building world, implementing these strategies at the design stages would allow more assurance on security and privacy for its users with enhanced service experience.

Though the metaverse is produced as a panacea for future digital expansion, there exist challenges and pragmatic issues hovering around it. Most critical issue being the lack of a serviceable digital infrastructure to offer the guaranteed services and applications with attributed processing and networking capabilities. Even such an infrastructure exists, access technologies required to offer the envisaged specifications are only viable with emerging 5G mobile technology, which is still in its experimental stages and not deployed globally. The compatibility and inter-operability between the virtual and physical worlds are to be understood and standardized before launching the metaverse. It is obvious that, even with a formidable level of a processing capability at the metaverse engines, resources might not be sufficient to serve the demand considering its potential and scalability with the social-media backbone. Thus, optimum processing and operation strategies must be adopted to alleviate the cost in terms of processing, storage, networking, and financial. Such strategies are only possible through automated AI based approaches and requires more assimilation and research on that subject. As a person is required to be equipped with a headset or a AR glass to access the metaverse at the minimum, the higher personal investment makes it a privileged service rather than a open system for everyone. Further, security and privacy of the users are imperative aspects, where certain privacy laws possible in the real-world might not be accountable in the virtual domain, while the prominent biometrics in the real-world can be replicated within the digital domain. Therefore, deploying the metaverse in a pragmatic context requires much more research and proper standardization.

XR is an obvious technology required for the metaverse development, where current AR and MR technologies should be improved to the level in which to advocate full integration of virtual entities to a super-realistic level, and to improve its omnipresence. AI, as specified earlier, plays a key role in automating the metaverse eco-system to hand in the complete control to the digital governance. AI involvement will further assure the prosumers of their digital assets and content safety, that is bounded by their avatar. Existing computer vision processing should be empowered with AI integration to enhance the 3D image processing, while image/video/3D rendering technologies can be improved to accelerate the query processing of the visual and telemetry data. XAI practices should be employed at the design stages to ensure global compliance for compatibility. As the existing cloud computing based storage and processing infrastructure lacks the required networking capability to host the metaverse applications, edge computing is a nascent paradigm essential to launch it that enables elevated access capacity along with context and location aware features due to its proximate nature. Further, network slicing can be employed to organize and structure the metaverse application flow among the eight enablers presented in \cite{lee2021all}. 

Introduced with the bitcoin cryptocurrency, blockchain rose to fame due to its unique ability to form a shared economy and laid the foundation to existing digital currency market. Blockchain has been considered a breakthrough technique for security and privacy preservation \cite{gadekallu2021blockchain}. In simpler terms, blockchain is a ledger that stores the committed transactions to facilitate digital asset tracing and securing in a commercial network. These transactions or records being stored as blocks, are linked together using cryptographic measures, or hashing mechanisms to be precise, ensures the ledgers` immutability and enables secure sharing capability even in an insecure environment. Most salient feature of blockchain is its capability to operate on decentralized ledger content without a centralized authority \cite{ynag2022fusing}. Since blockchain is employing proof of work as the consensus mechanism, the method itself deems it more secure and suitable for e-commerce platforms. In the context of the metaverse, blockchain is the pertinent enabler intended to enforce accountability into the digital eco-system.

The requirement for blockchain is imminent, where securing the digital content in possession of all the users of the metaverse is its prime purpose. The metaverse eco-system relies on blockchain for accounting their content and transactions to ensure user integrity, privacy, and reputation. 



\subsection{Related Works and Contributions}

The metaverse related studies and surveys are widely available and has incremented over the past few years \cite{ning2021survey}. In spite of its evident potential, blockchain based studies are limited and its wide adoption for various applications are not available at present. The authors in \cite{lee2021all} presented a comprehensive survey, and can be considered as the first scientific publication that discussed about the metaverse on a broader technical context. Blockchain has been introduced as one of the technological enablers out of seven other pillars. The authors identified data storage, data sharing, and data interoperability as the main uses of blockchain. A deeper study was not conducted on blockchain as the scope of the paper is quite wide. The important position blockchain holds in the metaverse inception is discussed in \cite{ning2021survey} concerning governmental and economic sectors while its possible utilization for virtual reality object connection is specified. However, this survey fails to build on those specified facts descriptively.

The idea of fusing blockchain and AI for the metaverse development was presented in \cite{ynag2022fusing} as a survey. This paper discusses the potential correlation between the metaverse and blockchain through the layered architecture composed of data, network, consensus, incentive, contract, and application layers. Though the authors present four blockchain empowered applications, they are mostly focused on the commercial usage of blockchain. Contribution of blockchain and AI for the metaverse was presented in \cite{jeon2022blockchain}, where handling and reusing high quality/ rich data, stabilizing the decentralized network, privacy of data, and handling of economic related data are discussed briefly. In addition, there are various studies \cite{mystakidis2022metaverse,wang2022metasocieties,park2022metaverse} that mention blockchain as a requisite for the metaverse, though fails to discuss them rigorously. Table \ref{tab:RWCompare} emphasizes the contribution of this paper. To the best of our knowledge, there hasn't been a study that discusses the utilization of blockchain for the metaverse applications. Thus, our study presents diverse potential applications for the metaverse where blockchain integration would enhance their efficiency, and impact of blockchain for enabling technologies.

The main contributions of this survey are:

\begin{itemize}
    \item Firstly, we present a brief overview of blockchain and the  metaverse, followed by the motivation behind integration of blockchain in the metaverse.
    \item Secondly, we discuss application of blockchain for addressing the challenges faced by several technical aspects of the metaverse including, data acquisition, data storage, data sharing, data interoperability, and data privacy preservation.
    \item Thirdly, we discuss about the impact of blockchain on some of the key enabling technologies in the metaverse such as Internet of Things, digital twins, multi-sensory XR and hologrtaphic telepresence, AI, and the big data.
    \item Fourthly, we discuss about some of the interesting projects such as Decentraland, Sandbox, Axie Infinity, and
Illuvium that leverage blockchain in the metaverse.
    \item Finally, we conclude the paper with some potential future research directions.
\end{itemize}

\begin{table}[!t]
\centering
    \caption{Emphasizing the contribution of this paper in contrast to the state-of-the-art}
    \label{tab:RWCompare}
\renewcommand{\arraystretch}{1.2}     
\begin{tabular}
  {|p{3.5cm}|c|c|c|c|c|c|c|c|}  
  \rowcolor{gray!25}
 
  \hline
   \cellcolor{gray!15} Context &{\rotatebox[origin=c]{90}{\cellcolor{gray!15}~\cite{ning2021survey}~}}
   &{\rotatebox[origin=c]{90}{\cellcolor{gray!15}~\cite{lee2021all}~}}
   &{\rotatebox[origin=c]{90}{\cellcolor{gray!15}~\cite{ynag2022fusing}~}}
   &{\rotatebox[origin=c]{90}{\cellcolor{gray!15}~\cite{jeon2022blockchain}~}}
   &{\rotatebox[origin=c]{90}{\cellcolor{gray!15}~\cite{mystakidis2022metaverse}~}}
   &{\rotatebox[origin=c]{90}{\cellcolor{gray!15}~\cite{park2022metaverse}~}}
   &{\rotatebox[origin=c]{90}{\cellcolor{gray!15}~Ours~}}
    \\
\hline   
\hline
   Metaverse Technical Perspective & \cellcolor{green!15}\checkmark & \cellcolor{red!15}\checkmark & \cellcolor{red!15}\checkmark & \cellcolor{yellow!15}\checkmark & \cellcolor{red!15}\checkmark & \cellcolor{red!15}\checkmark & \cellcolor{red!15}\checkmark \\
\hline
   Blockchain as an Enabling Technology for Metaverse & \cellcolor{yellow!15}\checkmark & \cellcolor{red!15}\checkmark & \cellcolor{red!15}\checkmark & \cellcolor{red!15}\checkmark & \cellcolor{green!15}\checkmark & \cellcolor{green!15}\checkmark & \cellcolor{red!15}\checkmark\\
\hline
   Applications of Blockchain & \cellcolor{yellow!15}\checkmark & \cellcolor{green!15}\checkmark & \cellcolor{yellow!15}\checkmark & \cellcolor{yellow!15}\checkmark &  &  & \cellcolor{red!15}\checkmark \\
\hline
   Technical Perspective of Adoptable Blockchain Methods & & \cellcolor{green!15}\checkmark & \cellcolor{yellow!15}\checkmark & \cellcolor{green!15}\checkmark & & & \cellcolor{red!15}\checkmark\\
\hline
   Impact of Blockchain for Metaverse Enablers  & & &  &  &  & & \cellcolor{red!15}\checkmark\\
\hline
 
\end{tabular}
\begin{flushleft}
\begin{center}
    
\begin{tikzpicture}

\node (rect) at (0,2) [draw,thick,minimum width=0.6cm,minimum height=0.4cm, fill= red!15, label=0:Technical Analysis] {\checkmark};
\node (rect) at (3.5,2) [draw,thick,minimum width=0.6cm,minimum height=0.4cm, fill= yellow!15, label=0:High Level] {\checkmark};
\node (rect) at (6,2) [draw,thick,minimum width=0.6cm,minimum height=0.4cm, fill= green!15, label=0:Introducing] {\checkmark};
\end{tikzpicture}
\end{center}

\end{flushleft}
  
\end{table}

\subsection{Paper Organization}
The rest of this paper is organized as follows. The fundamentals of blockchain,the metaverse, and the role of blockchain in the metaverse are presented in Section~\ref{Sec:Preliminaries}. The applications of blockchain for the metaverse from technical perspectives are discussed in Section~\ref{Sec:Technical}. In Section~\ref{Sec:EnablingTechnologies}, we discuss the impact of blockchain for enabling technologies in the metaverse, including IoT, digital twin, multi-sensory XR applications and holographic telepresence, AI, and big data. After that, we discuss about some of the exciting projects related to blockchain enabled metavserse applications in Section~\ref{sec:Project}. 
Finally, we conclude the paper with some potential research directions in Section~\ref{Sec:Conclusion}.

\section{Blockchain and The Metaverse: Preliminaries}
\label{Sec:Preliminaries}

The preliminary introduction to blockchain and the metaverse is presented in this section, followed by the role of blockchain in the metaverse. 
\subsection{Preliminaries of Blockchain}
\begin{figure}[!ht]
\centering 
\includegraphics[width=1.0\linewidth]{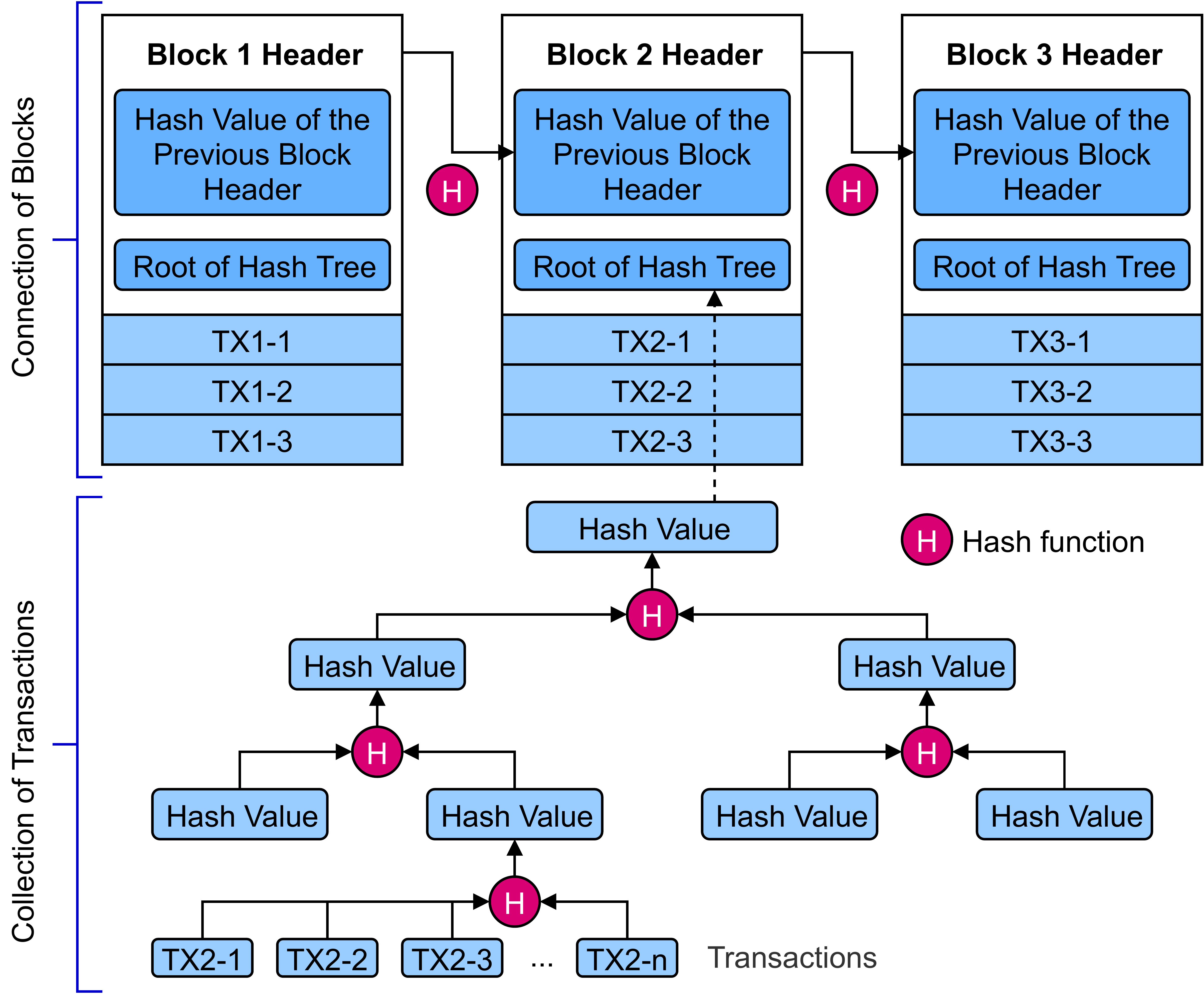}  
\caption{General structure of a blockchain, in which blocks connected with each other through their respective hash codes.}    
\label{fig:Blockchain framework}   
\end{figure}

\begin{figure}[!ht]
\centering 
\includegraphics[width=1.0\linewidth]{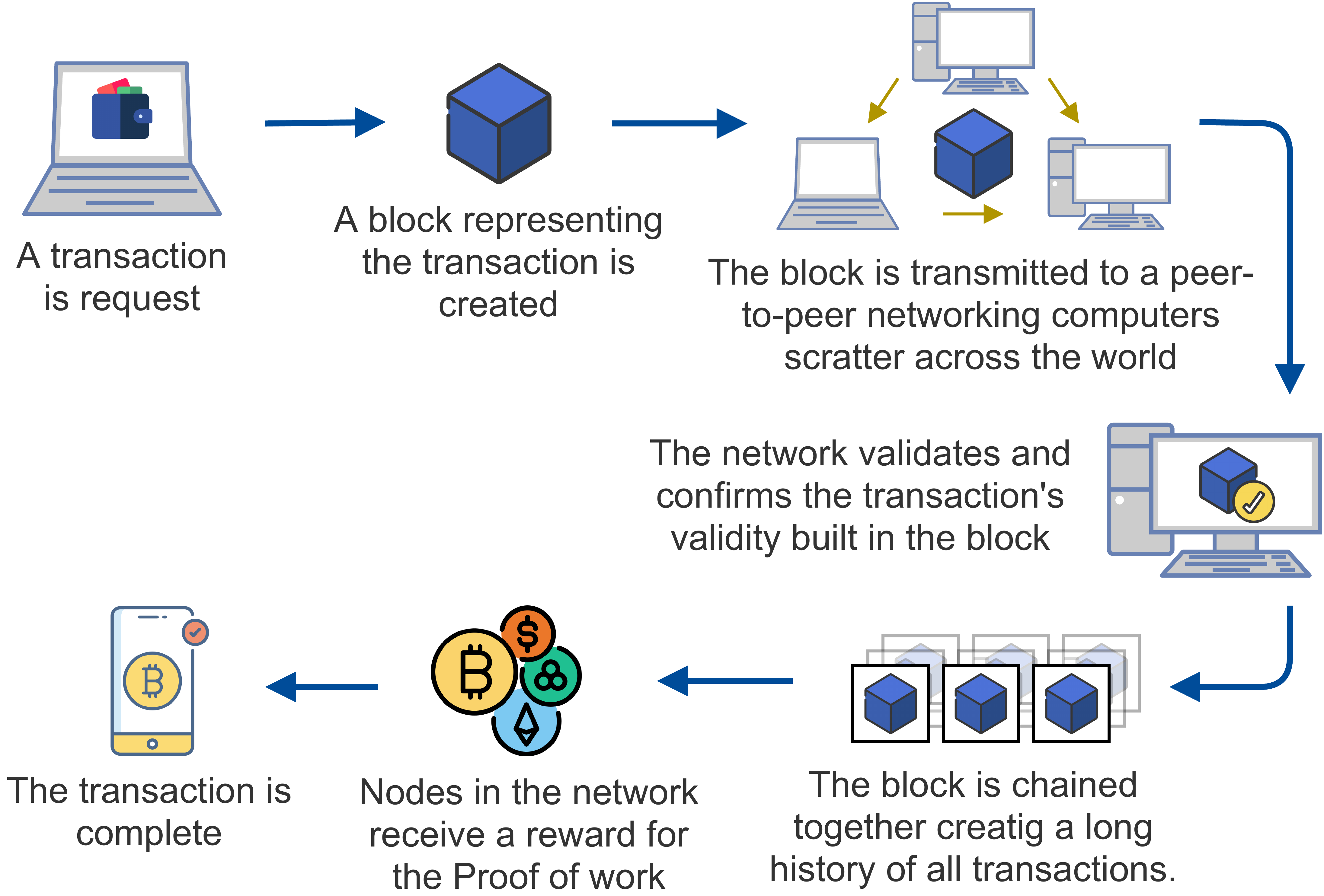}  
\caption{Illustration of a transaction processed by the blockchain technology.}    
\label{fig:how blockchain work}   
\end{figure}

The conception of blockchain origins from a white paper  written by Nakamoto Satoshi in 2008 \cite{nakamoto2008bitcoin}. Blockchain, also called distributed ledger, owns consecutive blocks, which are linked with each  other through the hash value of previous block header. Other than the inevitable cryptographic hash, timestamp, nonce and transaction data are also included in a block \cite{huo2022comprehensive}. The block timestamp is considered valid only if its value is more than the network-adjusted time plus two hours and greater than the median timestamp of prior eleven blocks, which prevents adversary to manipulate the blockchain possibly. Note that network-adjusted time refers to median of the timestamps from entire connected nodes. The smooth execution of blockchain is not just maintained by one or several nodes, instead, each node in the blockchain network should comply with a common consensus protocol to generate and validate new blocks. Consensus protocol is the backbone of blockchain where the operating principles and legitimate actions are all regulated \cite{dotan2021survey}. The famous bitcoin adopts Proof of Work (PoW) mechanism, which demands miners to contribute a great number of computing power to figure out an answer for the random mathematical problem \cite{alangot2021decentralized}. In order to avoid centralization of computing power, the difficulty, also called nonce of next block generation, is dynamically changed on the basis of 10 minutes per block.  Although unimaginable computation power hinder majority of attackers, PoW also leads to inefficient transaction rate and excessive energy consumption. Proof-of-Stake (PoS) alleviates the problems brought by PoW, so the miner who becomes ultimate winner depends on their quantity of holdings in the corresponding cryptocurrency rather than computing power \cite{thomsen2021formalizing}. The recent emerging InterPlanetary File System (IPFS) propagates Proof of Space (PoSpace) consensus, which requires participants to supply some storage space to prove a challenge posted by the service provider \cite{jian2021blockchain}. The transaction data is organized in the form of Merkle tree for each block, which improves the verification efficiency. Note that Merkle tree enables users to download any branch for check without full transaction records. The above-mentioned general parts of blockchain and the processing of transactions in a blockchain are depicted in Fig. \ref{fig:Blockchain framework} and Fig. \ref{fig:how blockchain work} respectively. 
 
 The representative of first generation blockchain is bitcoin, which only decentralizes the transaction records. Later, researchers found the blockchain can overload more functions such as asset management and family trust not merely a ledger. Hence, the second generation of blockchain-Ethereum came into the picture. The main innovation brought by Ethereum is the emergence of smart contracts \cite{zarir2021developing}. The rules of smart contract written into codes are stored in the blockchain. The transactions can be used to trigger the corresponding functions in the smart contracts, which automatically transfer funds or send notifications to the predetermined accounts. Since the smart contract is accessible to everyone, a series of applications spring up in a short time. For example, smart contracts strengthen the security of voting process, which becomes hard to be manipulated and decoded by the malicious guys. Moreover, the insurance industry unites hospitals to track and record patient information in the blockchain, where the smart contract can help the corporations claim settlement to the patient immediately. Recently, one of the smart contract derivation--Non-fungible tokens (NFTs) are popular around the world. Ethereum issues some standards such as ERC-721 and ERC-1155 to introduce the features of assets built on NFTs. Unlike cyrptocurrencies in Bitcoin and Ethereum, each NFTs are non-interchangeable and cannot be divided. The NFT market is now worth more than \$7 billion dollar, including 
art, games, sports, copyright, insurance and many other areas \cite{nadini2021mapping}. 


\subsection{Preliminaries of the  Metaverse}

    \subsubsection{What is the metaverse?}
  The word ``meta'' origins from the Greek language, which is a prefix that means ``more comprehensive'' or ``transcending''\footnote{https://en.wikipedia.org/wiki/Met}. The word ``Verse'' is abbreviated version of universe, which represents a space/time container \footnote{https://alldimensions.fandom.com/wiki/Category:Verse}. When these two words are combined, a brand new word ``Metaverse'' comes into the picture, where the traditional social systems are transformed into a novel digital living space. State-of-the-art technologies such as virtual reality (VR), digital twin, blockchain, etc are used to build the metaverse that maps everything in our real world \cite{wang2022metasocieties} to a parallel universe. For example, users can work, live and play with friends from any place in the metaverse. In 1992, Neal Stephenson, in his famous science fiction novel called Snow Crash \cite{stephenson2003snow} firstly proposed the initial conception of the metaverse, where people utilize digital avatars to control and compete with each other to upgrade their status. However, till now, the metaverse is in its conceptional stage; there are common standards and very few real implementations are available.

    \subsubsection{What are enabling technologies of metaverse?}
The metaverse is a fusion of multiple emerging technologies such as 6G, artificial intelligence (AI), VR and digital twins. The core technologies required in the metaverse are:
\begin{enumerate}
    \item The most important technology for realization of the metaverse is extended reality technology, including AR and VR. While AR can overlay and superimpose digital information onto the physical environment, VR allows users to experience the digital world in a vivid way \cite{koutitas2021performance}. Both these techniques are very important in the development of the metaverse, which creates digital space where the users can interact as in the real world.
    \item The second important technology is digital twin, which establishes a virtual twin of a real world object by utilizing real world data to predict the expected behavior of the real world object \cite{tao2018digital}. In the metaverse, digital twin can mirror the real world into the virtual world. Correspondingly, the metaverse can also find some trial solutions to the unsolved issues in the real world.
    \item The third technology--blockchain plays two irreplaceable roles in the metaverse. On one side, blockchain technology serves as a repository, so users can use it to store data anywhere in the metaverse. On the other side, blockchain technology can provide a complete economic system to connect the virtual world of the metaverse with the real world. Especially, the above-mentioned NFTs allow virtual goods to become physical objects. Users are allowed to trade virtual items in the same way as in the real world. Hence, blockchain bridges the real world and the metaverse \cite{jeon2022blockchain}.
\end{enumerate}

    \subsubsection{ What are applications of the metaverse?}
    Some of the popular applications of the metaverse are as follows.
    \begin{itemize}
    \item \textbf{Online video conference}: In the surreal atmosphere of the Covid-19 pandemic, many small corporations are kept alive through the application of telecommuting. However, as we all know, face-to-face communication is significant, 70\% expression of people comes from body language rather than verbal language.  Telecommuting has a number of problems unlike traditional face-to-face collaboration, such as inefficient cooperation, delay interaction and misunderstood feedback \cite{chang2021telecommuting}. However, in the metaverse, people can utilize a friendly avatar to walk around and work in the virtual space. Even the body language or eye interaction can be used to communicate with working partner from different angles, which will greatly improve telecommuting.

    \item \textbf{Digital Real Estate}: In general, real estate refers to property consisting of land and buildings, which can be used for establishment, living, investment, rent, sale and buy. In the metaverse, the above-mentioned activities can also be implemented. Besides, the same factors in the reality such as location, amenities and transportation can influence house prices in the virtual world. For example, in the metaverse, users can collect and sell households to the public for a second time, as well as organize art exhibitions, music festivals, gaming competitions and so on. The virtual platform explicitly emphasises the scarcity of virtual land to users, which is offered to users at auction and traded for NFTs \cite{duan2021metaverse}.

    \item \textbf{Digital Arts}: Traditionally, people establish the 3D images through some modelling tools such as Maya and ZBrush. However, the metaverse has a strong focus on the display layer, which brings new ways of expression and artwork creation, so we can draw figures by using a brush directly. On one side, with the growing popularity of AI painting, digital art is gradually coming to the public's attention. On the other hand, the emerging blockchain technology has also brought traditional artworks from offline to online. In the virtual gallery placed in the metaverse, the users can enter the gallery to appreciate from all dimensions \cite{ko2021study}.
\end{itemize}

\subsection{Role of Blockchain in the Metaverse}
When it comes to the metaverse, the imagination of a variety of dazzling experiences or fun games may rise in our mind. However, the scenario closely related to us is just a parallel world, where the economic ecology is inevitable. Moreover, the digital assets are the core functions provided by the blockchain, such as the homogenized tokens based on ERC-20 and the non-homogenized tokens based on ERC-721 or ERC-1155. Since the blockchain technology can maintain the smooth economic operation of metaverse, blockchain technology is the soul of the metaverse. 

The motivation behind integration of blockchain in the metaverse are summarized below
\begin{itemize}
    \item \textbf{Ensuring Data Privacy and Security}:
    The metaverse collects vast volumes of sensitive information in order to present the user with the greatest possible experiences. The organizations or the applications need this data for the successful development of targeting systems. If the information is leaked into the hands of the wrong people they might also target users in the real world. Blockchain, with its authentication, access control, and the consensus mechanisms provide the users complete control of their data thereby securing the data privacy of the users. The blockchain uses asymmetric-key encryption and hash functions which ensure security of data in the metaverse.      
    
    \item \textbf{Ensuring the Quality of the Data}:
    The metaverse receives data from multiple applications ranging from healthcare to entertainment. The AI models in the metaverse rely on this data for making key decisions for its stakeholders. The creation of the objects in the metaverse relies highly on the quality of data shared by the users from the real world. Blockchain, provides complete audit trails of transactions, allowing individuals and organizations to validate all transactions. This will increase the data quality in the metaverse.    
    
    \item \textbf{Enabling Seamless and Secured Data Sharing}:
    The metaverse depends on AR and VR devices, resulting in a more connected and immersive world. The metaverse's real benefit resides in its integration with AR on digital and physical objects. The metaverse's success is dependent on the seamless sharing of AR and VR data, which enables the development of new, advanced applications that aid in resolving real-world problems. The blockchain's advanced encoding information system enables the metaverse's data sharing to be seamless and secure.
    
    \item \textbf{Enabling Data Interoperability}:
    In the metaverse, stakeholders need to access and hold assets in different virtual worlds and use a variety of applications. Data interoperability across these virtual worlds is limited due to the different environments in which they are built. It is possible to exchange data on two or more blockchains located in distinct virtual worlds using a cross-chain protocol. Users can migrate more easily between these virtual worlds because of the blockchain's interoperability.
    
    \item \textbf{Ensuring Data Integrity}:
    The metaverse's data must be maintained consistently and accurately. The stakeholders may lose faith in the metaverse if the integrity of the data is compromised. The metaverse data is saved as a copy in every block throughout the chain that can't be amended or removed without the consent of majority of the participants, due to the immutability provided by the blockchain. This mechanism of blockchain ensures the data integrity of the metaverse.
\end{itemize}

Some use cases of application of blockchain in the metaverse are discussed below.
\begin{itemize}
    \item \textbf{Financial system}:
    Tamper-proof, openness, transparency and decentralization are the four significant features in blockchain. In the metaverse, millions of transactions happen for goods exchange in a short time \cite{kim2021digital}, so the security and efficiency of these transactions must be guaranteed. Based on the above-mentioned features, blockchain is a good candidate for the large-scale and scalable economic system construction in the virtual world.

    \item \textbf{Smart contract deployment}:
    The inherent nature of the blockchain network allows smart contracts to be automated, programmable, open, transparent and verifiable among other remarkable features, thus allowing for on-chain trusted interactions without the need for a third-party verification platform. If the financial system in the metaverse is built on top of the blockchain, the characteristics of smart contracts can be used to decentralise the operation of contracts in a programmed, non-custodial, verifiable, traceable and trustworthy manner, thus significantly reducing the harmful behaviours such as rent-seeking, corruption and underhanded operations that may exist in the financial system, and can be widely used in the financial, social and gaming sectors.
    
    \item \textbf{NFTs}: 
    The most important feature of NFT is indivision and uniqueness, which make it suitable for identity representation, for example, assets that are exclusive and indivisible and can be freely traded and transferred. In the metaverse, these virtual assets come with certificates called NFTs that indicate ownership \cite{nadini2021mapping}.
\end{itemize}

\section{Blockchain for the Metaverse: Technical Perspective}
\label{Sec:Technical}

\begin{figure*}[!ht]
\centering  
\includegraphics[width=1.00\linewidth]{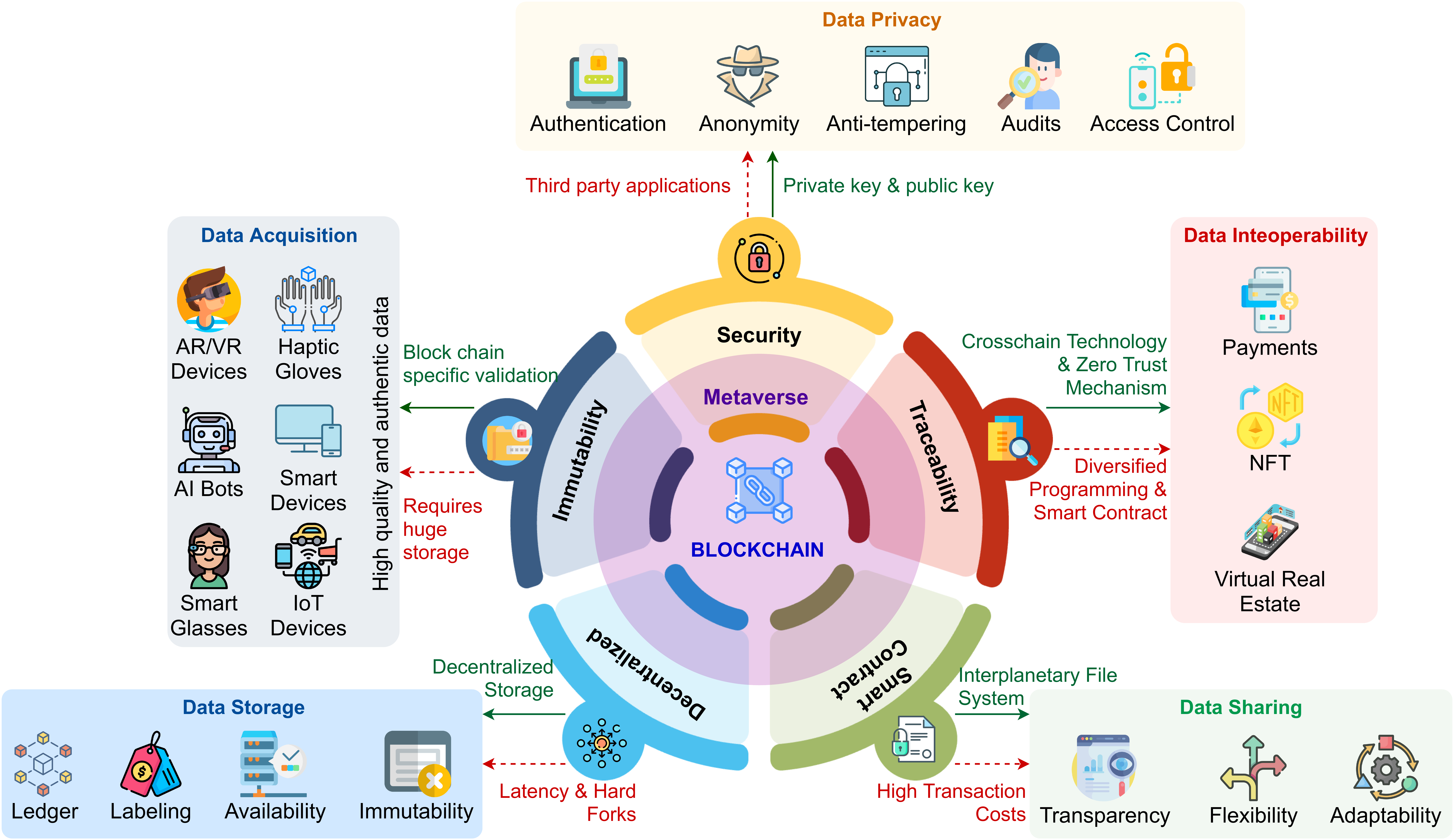} 
\caption{Blockchain for technical aspects in the metaverse.}    
\label{fig:Bc_tech}    
\end{figure*}

This section investigates the state-of-the-art blockchain-based methods for the metaverse from the technical perspectives, including data acquisition, data storage, data sharing, data interoperability, and data privacy preservation. The illustration of blockchain for the above-mentioned technical aspects in the metaverse is depicted in Fig.~\ref{fig:Bc_tech}.


\subsection{Data Acquisition}
\subsubsection{Introduction}

Data acquisition is an important step in the metaverse ecosystem. Some of the sensitive data from the users such as bank/credit card details are acquired when the users make payments. In addition, sensitive data such as bio-metrics poses/gestures of the users have to be acquired in the metaverse to create digital avatars \cite{cha2021performance}. Data acquisition helps in training the AI/ML algorithms that can assist in decision making, digital product development, recommendation system development, and marketing in the metaverse \cite{wang2022metasocieties}. 
Data acquisition in the metaverse will help applications create better insights, able to change over time, and adapt to new situations. The metaverse will be a digital marketplace where people can buy, sell, play, talk, and work digitally using various devices as depicted in Fig.~\ref{fig:Bc_tech}. Hence, massive amounts of heterogeneous data will be generated \cite{xi2022challenges}. Some of the data acquisition methods that can be used in the metaverse are as follows. Web forms will be one of the data collection tools in the metaverse for gathering information from users. Clients can quickly and easily fill out information using web forms before being granted access to the metaverse capabilities. Bots can capture personal information and required authentication elements like the social security number from the users \cite{jovanovic2022vortex}. A high-definition camera in the metaverse helps to gather information about a user's physical attributes. The physical attributes of the user are used to build a digital representation of the user in the virtual world \cite{duan2021metaverse}. The client will benefit from the use of AR/VR gadgets to explore the metaverse. These gadgets will be used to gather information about the user's behavior \cite{rauschnabel2022augmented}.

\subsubsection{Challenges of Data Acquisition in the Metaverse}

Data generated through decentralized applications (Dapps) like  WeiFund, Etheria, 4G Capital, and Ampliative Art in the metaverse will be huge, unstructured, and real-time. It poses a significant challenge in acquiring the enormous data being generated. Data assurity or integrity will become important for building applications like recommender systems in the metaverse. These systems will be affected if the data is gathered from unknown sources as it can impact the reliability of these systems \cite{tao2018secured}. The volume of data will skyrocket in the metaverse as it provides high-quality digital services like Horizon Worlds and Horizon Venues applications using VR headsets \cite{brunschwig2021towards}. This will create a burden on data acquisition systems due to increase in streaming in entertainment and other applications \cite{jeong2022rethinking}. Duplicate and inaccurate data may also be acquired which will affect the quality of data \cite{shiau2022scale}.

\subsubsection{How Blockchain Can Help}
Acquisition of authentic data in the metaverse will be made easier for applications like social networking with the adaption of blockchain technology. Distributed ledger in blockchain will allow validation of transaction records and tracing the data in the metaverse \cite{islam2019buav,deepa2022survey}. As a result, data acquisition is resistant to attacks as the majority of nodes in the ledger must approve any changes to the data in the metaverse \cite{xu2021light}. All the data acquired in the metaverse is subjected to a blockchain-specific validation procedure which is powered by consensus mechanisms \cite{bouraga2021taxonomy,lashkari2021comprehensive}. In a blockchain, every activity is recorded as a transaction and each block contains a cryptographic hash of the previous block along with a timestamp and the metadata \cite{luo2021novel}. Hence, in a block, data cannot be altered without altering the other blocks. The data obtained from any block is resistant to tampering \cite{zhang2021research}. The chance of creating a duplicate block is almost zero which ensures no duplication in the process of data acquisition. As every block is authorized in the blockchain, the data acquired through blockchain enabled acquisition systems in the metaverse will be reliable \cite{guo2021reliable}.

\subsubsection{Summary}
In the metaverse, data acquisition presents a challenge in terms of ensuring high-quality and authentic data. Although blockchain technology will enable data acquisition systems to overcome these constraints,  blockchain can be slow due to its complexity and distributed nature \cite{xu2021latency}. Transactions on a blockchain can take much longer to complete.  The entire transaction may take a few days. As a result, transaction fees are higher than usual, and the number of users on the network is limited \cite{alrubei2020latency}. Data collected in a blockchain must be copied along the chain, increasing storage demand. The more data collected, the more storage space is needed \cite{chen2022blockchain}. To address these issues in data acquisition systems, there is still room for research in developing matured blockchain for the metaverse.

\subsection{Data Storage}
\subsubsection{Introduction}
The metaverse is a digital realm that exists alongside the physical world and is governed by humans. The metaverse will comprise of experiences, places, and the things accessible over the Internet. 
The metaverse requires massive amounts of data storage. Every person who enters the metaverse creates a data file, and the data continues to grow as a result of social interactions. Massive amounts of data will be generated once the metaverse is built and implemented, laying a significant strain on the real world's ability to process that information. Data storage will have to be a top priority in order to put the metaverse to use \cite{duan2021metaverse}. 

\subsubsection{Challenges of Data Storage in the Metaverse}
In the metaverse, a digital reality exists alongside the physical world. As more people join the digital worlds, large amounts of data files are created, and as a result, the metaverse will generate voluminous data. The physical world's data storage capacity will be pushed to the limits as soon as the metaverse is fully operational. As a result, data storage will be a major challenge for deploying metaverse applications like gaming, entertainment, real estate, healthcare, etc \cite{bian2021demystifying}. There is a risk of data leakage, tampering, or loss if the metaverse relies on a central storage system. The high probability of data loss  \cite{wang2021joint} and corruption in centralized applications endangers the metaverse's ability to provide biometric data, vocal inflections, and vital signs that rely on sensitive data \cite{kiong2022metaverse}. Data labeling and organization will be another significant challenge with huge amounts of data produced by the metaverse applications \cite{scargill2022environmental}.

\subsubsection{How blockchain can help}
For every transaction, a new block is created, making the metaverse storage impenetrable to tampering \cite{liang2020secure}. Consequently, data is saved as a replica of the original blocks throughout the chain, boosting data reliability and transparency in the metaverse \cite{jeon2022blockchain}. The metaverse applications, ranging from real estate to digital objects will be at high risk if the centralized data storage is compromised \cite{yang2022expert}. The use of blockchain technology will result in numerous blocks contributing to data distribution and thereby increasing data availability in applications like vital monitoring and life support alerts in the metaverse. The decentralized nature of blockchain technology allows data scientists in the metaverse to collaborate and work on data cleansing, which will significantly reduce the time and costs associated with labeling data and preparing datasets for analytics \cite{label}.

\subsubsection{Summary}
The decentralized nature of blockchain technology will reduce the time it takes to identify and label data while also serving as a collaborative platform for data scientists. Furthermore, in the metaverse, the blockchain provides data reliability, transparency, and availability as depicted in the Fig.~\ref{fig:Bc_tech}. The data will be backed up in every block of the blockchain. A consensus-based distributed ledger will help the data in the metaverse to be resistant to tampering and duplication \cite{xie2019survey}. However, more research is required to address the issue of latency, as any data added must be mirrored throughout the entire chain. Although data tampering is impossible in blockchain, a hard fork is a possibility that must be considered.


\subsection{Data Sharing}
\subsubsection{Introduction}
Data sharing can benefit a diverse spectrum of the metaverse stakeholders in numerous ways. As people and applications share the same platform, they may collaborate more effectively as depicted in Fig.~\ref{fig:Bc_tech}. Everyone, from scientists to the general public, will benefit from data exchange in the metaverse \cite{kraus2022facebook}. The data collected from AR/VR and IoT devices in the metaverse will be used to create personalized systems that are customized to the users' actions. This will enable a wide variety of applications to deliver a more positive user experience \cite{jeon2021effects}. Organizations will be able to conduct data analytics through the metaverse by disseminating information across applications. Shared data will help understand customers, evaluate advertising, personalize content, establish content strategies, and build products in the metaverse \cite{kiong2022metaverse}.

\subsubsection{Challenges of Data Sharing in metaverse}
Sharing data in centralized data exchange platforms can expose sensitive and private data of the data owners to heavy risk in the metaverse \cite{liu2020blockchain,egliston2021critical}. Data in the traditional sharing environment is highly mutable, this results in high latency and lower data availability. Scaling the mutable data is challenging compared to the immutable data \cite{yu2021blockchain}. In the metaverse, numerous applications like healthcare, traffic optimization, media, and entertainment will generate large volumes of data and will be mostly real-time. Data flexibility becomes an issue when the demand for real-time data increases in a traditional data-sharing environment.

\subsubsection{How blockchain can help}
Blockchain technology can make the transactions in crypto exchange, education, and other applications more transparent and precise in the metaverse \cite{egliston2021critical}. Applications like governance and finance will generate a decentralized, immutable record of all transactions, allowing stakeholders to view these records. Hence, the stakeholders of the metaverse will benefit from greater data transparency \cite{rashid2021blockchain}. Blockchain will enable applications and their users to understand how third-party applications like thunderbird, the bat, and pegasus manage data and can eliminate grey market transactions which will boost user confidence \cite{vashistha2021echain}. In addition, the data owner will have complete control over the information. Data audits can also benefit from distributed ledger technology. As a result, blockchain reduces the time and money spent on validating the data \cite{min2022portrait}. Smart contracts will improve the flexibility in data sharing. They are typically used to automate the execution of an agreement so that all participants can be certain of the outcome immediately, without the involvement of an intermediary or the loss of time. Blockchain allows smart contracts to be heterogeneous in programming. Hence, it will benefit applications like Nmusik, Ascribe, Tracr, UBS, and Applicature \cite{ali2021comparative}.   

\subsubsection{Summary}
The use of blockchain technology will improve the flexibility and adaptability of the metaverse data. A blockchain must replicate copies of data along the chain resulting in greater delay when transferring information \cite{luo2019blockchain}. As the number of people in the metaverse increases, the number of blocks must increase as well, necessitating the use of massive amounts of computing resources \cite{gao2021b}. Users will be charged a higher transaction cost for the validation of shared transactions as a result of this. Future-generation blockchains have to address this issue for effective sharing of the data in the metaverse.

\subsection{Data Interoperability}
\subsubsection{Introduction}
Interoperability will be the major driving force behind the metaverse.  A diverse set of applications like finance and healthcare will be able to communicate and exchange information in the metaverse. The metaverse will be a social and cultural interaction platform for virtual worlds. virtual bridges will be created progressively to allow users to keep their avatars and possessions while easily transferring them between virtual worlds. A unique set of credentials is issued to the user using an identity standard, and these credentials can be used across virtual world borders \cite{sparkes2021metaverse,kiong2022metaverse,stokel2022welcome}. This could be the same as our real-life license numbers, social security numbers, passport numbers, and other identification numbers.

\subsubsection{Challenges of Data Interoperability in the Metaverse}
The metaverse will be created through the fusion of numerous digital realms. The traditional centralized digital platforms that are currently available are disjointed and unorganized. Individuals must set up their accounts, avatars, hardware, and payment infrastructure to participate in different realms \cite{bian2021demystifying}. A user's options for transferring his or her digital possessions like NFT and avatar to another digital environment are restricted. It is tough to relocate in virtual worlds due to lack of openness, for example, using the same account in decentraland to roblox is not possible. The potential to use an application in the virtual world will depend on the interconnection among the virtual worlds. Regardless of where they are located or what technology is being utilized, digital world applications should be able to freely communicate information with one another. The metaverse interoperability is dependent on the capacity to manage the interactions between virtual worlds in an appropriate manner which is a serious limitation of the traditional approach \cite{mystakidis2022metaverse}. 

\subsubsection{How blockchain can help}
To ensure interoperability between virtual worlds in the metaverse, a cross-chain protocol is a perfect solution \cite{belchior2021survey,madine2021appxchain}. This allows the exchange of possessions like avatars, NFT's, and payment between virtual worlds as depicted in the Fig.~\ref{fig:Bc_tech}. This protocol will provide the groundwork for widespread the metaverse adoption. Interoperability between virtual worlds will be enabled through the use of cross-blockchain technology
\cite{jabbar2020blockchain}, eliminating the need for intermediaries in the metaverse.  Blockchain will make it simple for people and applications to connect in the metaverse.

\subsubsection{Summary}
Despite the potential of blockchain in increasing the interoperability between virtual worlds in several metaverses, further research is required. The main challenge in cross-blockchain enabled the metaverse interoperability is the existence of several public blockchains in different virtual worlds that do not share a common language. Various platforms will provide varying degrees of smart contract capabilities, making adaptation difficult. Additionally, the transaction architecture and consensus processes utilized in these virtual worlds vary considerably, limiting interoperability \cite{wibowo2019improving}.

\subsection{Data Privacy Preservation}
\subsubsection{Introduction}
The metaverse will make use of advanced human-computer interface (HCI) technologies,  allowing users to participate in social interactions as well as interact with their virtual surroundings \cite{siyaev2021towards}. Web 2.0 is centralized and raises concerns about data privacy. As the Internet of the metaverse, also known as Web 3.0 grows in scope and complexity, the metaverse will reduce the boundaries between the real world and the virtual world \cite{arvas13gutenberg}. The consequences of the Web 2.0 impact on personal rights protection are yet to be addressed. Hence, the problem of data privacy will become even more complicated in upcoming Web 3.0. The amount of data generated in the metaverse will increase dramatically; hence, inadequate security procedures of applications will raise the likelihood of a data breach \cite{kostenko2022electronic}. Privacy preservation is a major cause of concern when it comes to safeguarding the confidentiality of personal data \cite{jovanovic2022vortex}. 

\subsubsection{Challenges related to Privacy Preservation in the metaverse}
The metaverse ecosystem will be difficult to adapt at the initial stages where the attackers can trick the users and can steal sensitive information. If an artificial intelligence bot like promobot is deployed, the user will not be aware of whom they are dealing with, and the user may believe that they are interacting with a real person, resulting in them being deceived. Personal identifiable information (PII) is a cause of concern when it comes to safeguarding the confidentiality of personal data \cite{hughes2022metaverse}. Integrating validity information in the metaverse will increase the difficulty to manage large volumes of the data at the same time.

\subsubsection{How blockchain can help}
Blockchain technology gives users of the metaverse the ability to control their data through the use of private and public keys, effectively granting them ownership of their data. In the blockchain-enabled metaverse, third-party intermediaries are not permitted to misuse or gain data from other parties. In the case of personal data kept on the blockchain-enabled the metaverse, the data owners will be able to regulate when and how a third party can access their information \cite{kumar2021ppsf}. An audit trail is included as a standard in blockchain ledgers, ensuring that transactions in the metaverse are complete and consistent  as depicted in the Fig.~\ref{fig:Bc_tech}. The adoption of zero-knowledge proof on the blockchain allows individuals with convenient access to the identification of essential data in the metaverse while protecting their privacy and maintaining ownership over their possessions. Zero-knowledge proofs are a mechanism of blockchain by which users can persuade applications that something about them is real without disclosing the information \cite{sedlmeir2021next}.

\subsubsection{Summary}
The adoption of blockchain technology can assist users in the privacy preservation of their data; but, a single human error, such as the loss of a private key, has the potential to compromise the security of blockchain technology and the privacy of data in the metaverse. In the metaverse, attackers can easily target third-party applications since they tend to make use of inadequate security mechanisms, resulting in the compromise of personal information \cite{hassan2019privacy}. There is still a lot of potential for an investigation into how blockchain can be used in the metaverse to ensure user data privacy.

\section{Impact of Blockchain on Key Enabling Technologies in the Metaverse}
\label{Sec:EnablingTechnologies}



\begin{figure*}[ht!]
\Centering 
\includegraphics[ width=1\linewidth]{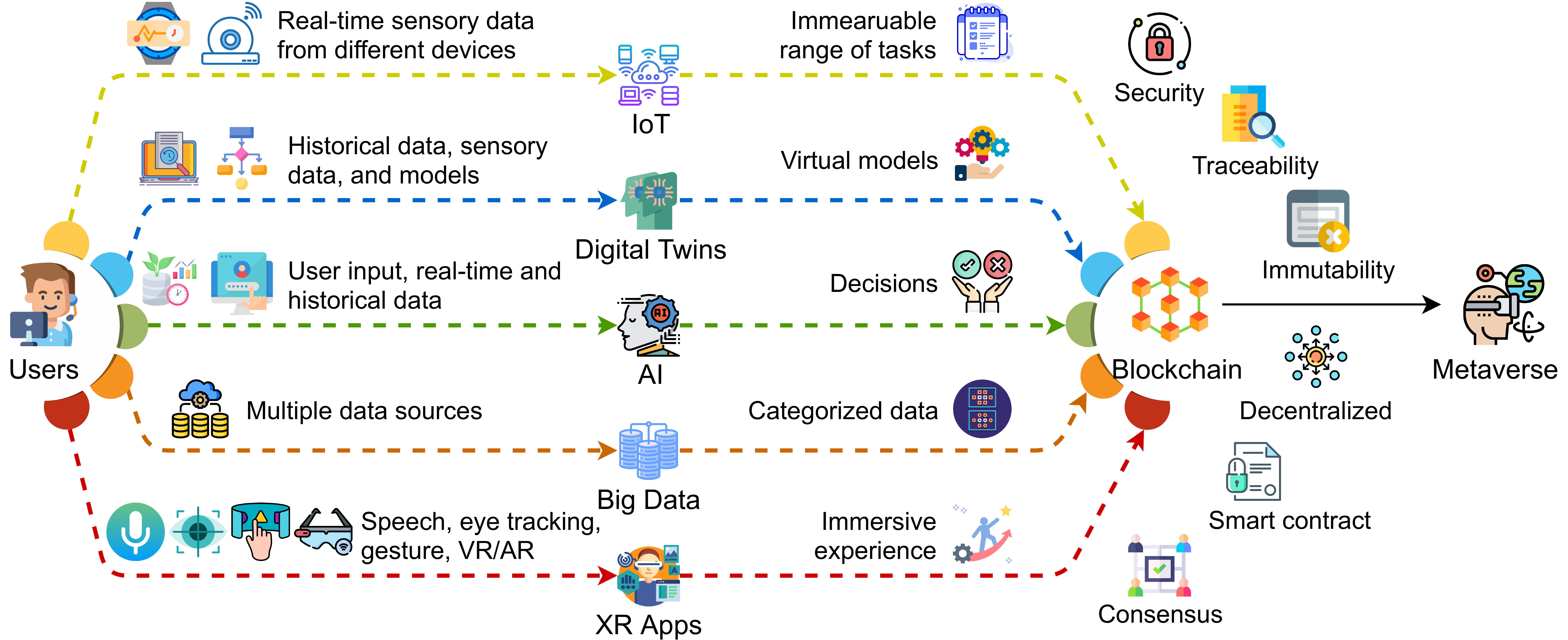}
\caption{Blockchain for key enabling technologies of the metaverse.}    
\label{fig:Bc_et}    
\end{figure*}

Blockchain can enhance the key enabling technologies in the metaverse that will allow users in participating social and economic activities without fear of repercussions. The illustration of blockchain on key enabling technologies in the metaverse is depicted in Fig.~\ref{fig:Bc_et}.


\subsection{Blockchain for IoT in the Metaverse}
\subsubsection{Introduction}

The metaverse platform gathers data from a variety of Internet of Things (IoT) devices to ensure that it runs efficiently in several applications of the metaverse such as medicine, education, and smart cities \cite{kanter2021metaverse}. The IoT devices will connect the metaverse through the use of a diverse range of hardware, controllers, and physical items. Connecting to the metaverse and navigating both physically and virtually is made possible by IoT devices equipped with specialized sensors. The capacity of IoT devices to perform operations in the metaverse will be critical to the user's ability to operate in the metaverse \cite{yang2022expert}.

\subsubsection{Challenges related to IoT in the Metaverse} 

There will be a huge number of connected IoT sensors in the metaverse. With so many connected devices, IoT storage and security are undoubtedly a concern. It is incredibly difficult to analyze IoT data that is unstructured and real-time \cite{zhang2021blockchain}. The quality of data can be judged by the amount, precision, and speed of data \cite{rollo2021air}. In addition, the metaverse data must be error-free for analysis. The use of a centralized strategy is not advantageous when it comes to storing data across virtual worlds. If even a single piece of data has been tampered with, it will harm the entire set of results produced by the IoT devices. The cross-platform capabilities of IoT devices are critical for sharing data between virtual worlds \cite{hajjaji2021big}. IoT data must be tracked for safety and regulatory compliance reasons.

\subsubsection{How blockchain can help}

Blockchain enables the IoT devices in the metaverse to communicate data through cross-chain networks, which in turn produce tamper-resistant records of shared transactions in virtual worlds as depicted in Fig.~\ref{fig:Bc_et} . As a result of blockchain technology, applications and users will be able to share and access IoT data without the need for centralized management or control \cite{majeed2021blockchain}. To eliminate conflicts and increase confidence among the metaverse users, each transaction is recorded and authenticated. In the metaverse, IoT-enabled blockchain enables the storage of data in real-time. All stakeholders can rely on the data and take action promptly and efficiently because of immutable blockchain transactions \cite{dorri2021temporary}. Blockchain technology allows stakeholders to keep track of their IoT data records in shared blockchain ledgers this will help resolve issues in the metaverse. 

\subsubsection{Summary}

Blockchain will enable IoT devices to share and store real data securely over multiple virtual worlds. Blockchain technologies require a significant amount of processing power to keep them running. Blockchains are vulnerable if a small group of miners controls most of the network's total mining hash rate. It is not possible to verify  IoT data that is not public due to a lack of governance before it is published on the blockchain in the metaverse. Smart contracts in the metaverse that are executed on a distributed transaction ledger may violate the laws. It is difficult to trace down all IoT transactions involving unlawful services in the metaverse because of the anonymity provided by blockchain technology. While the blockchain's automatic function offers numerous advantages, pinpointing which parties are responsible for specific behaviors remains a difficult challenge \cite{uddin2021survey}. The blockchain should be regularized to carry out the expansion of the metaverse.

\subsection{Blockchain for Digital Twins in the Metaverse}

\subsubsection{Introduction}

Digital twins \cite{ramu2022federated} are sophisticated digital representations of everything in the metaverse, ranging from simple assets to complex products and surroundings. Therefore, anything that is relevant to the user's needs could be a component of the ecosystem using digital twins. Two-way IoT connections enable users to bring their preferred models to life while keeping them in synchronization with the actual world. The applications of the metaverse will not be able to work properly unless a connection between the physical and digital worlds is established at the beginning. Digital twins will be important to understand how the metaverse environment will evolve and will aid in the prediction of the future \cite{chen2021research}. Using digital twins, it is possible to predict when hardware will need to be serviced or to estimate the demands of users before they arrive in the metaverse \cite{yoon2021interfacing}.

\subsubsection{Challenges related to Digital Twins in the Metaverse}

Digital twin models in the metaverse will be developed using information obtained from several remote sensors. Digital twin model accuracy is affected by the quality of the data that is used to create the model. In other words, data provided by the source must be genuine and up to standard in terms of quality \cite{zhuang2021digital}. Collaboration between digital twins in different virtual worlds should be possible. It will improve the outcomes of the metaverse. Digital twins need to interact and link to other digital twins ranging from healthcare to financial markets. Virtual worlds are constantly changing, and digital twins in the metaverse should detect and respond to these changes. The digital twins should be capable of identifying and fixing errors, which results in more accurate and consistent communication. When a variety of devices and sensors are brought together to develop digital twin models using real-time data, it is challenging to keep data safe from botnets and other malware \cite{khan2022digital}.

\subsubsection{How blockchain can help}

Blockchain encryption capabilities and historical data transparency enable digital twins resistant to attacks and securely share data \cite{lee2021integrated} over different virtual worlds. Data can be shared between digital twins in virtual worlds using an intelligent distributed ledger. Real-world objects will be stored on the blockchain and synchronized to digital twins in the metaverse using an intelligent distributed ledger as depicted in Fig.~\ref{fig:Bc_et}. Additionally, the deployment of digital twins on a blockchain will aid in the resolution of issues related to privacy and data security \cite{shen2021secure}. By merging blockchain with AI, it will be feasible to track sensor data and produce high-quality digital twins in the metaverse. Every digital twin action in the metaverse will be recorded as a transaction on the blockchain, which is immutable and requires consensus to change \cite{lee2021integrated}.

\subsubsection{Summary}
The incorporation of blockchain technology into digital twins enables the metaverse stakeholders to efficiently manage data on a shared distributed ledger while also addressing data trust, integrity, and safety concerns. Standardization, privacy, and scalability are all issues that must be addressed for blockchain to be successfully implemented in digital twin applications in the metaverse. The combination of blockchain, XAI, and federated learning approaches will improve the quality of digital twins in the metaverse \cite{wang2021explainable}.

\subsection{Blockchain for AI in the Metaverse}

\subsubsection{Introduction}

AI is one of the most important enabling technology for the foundation and development of the metaverse, which helps it in reaching its full potential. Based on the original image or 3D scan, an AI model will analyze user images automatically and create a very realistic simulated rendition called avatars. In the metaverse, the representational attributes and features of avatar affect the overall quality of user experience. Concretely, AI can plot a variety of facial expressions, feelings, fashions, aging-related characteristics, and so on for the avatar to make it more dynamic \cite{duan2021metaverse}. As a result of significant artificial intelligence training, the metaverse will be available to individuals all across the world, regardless of their linguistic competence. Making the metaverse experience that is both entertaining and authentic while salable will be challenging without the use of AI.

\subsubsection{Challenges related to AI in the Metaverse}

In terms of science and technology, the metaverse represents a new frontier, and establishing AI there will be a difficult task. It is difficult to track down the ownership of AI-powered material in the metaverse. Users have no means of knowing whether they are interacting with a real person or a computer-generated avatar as depicted in Fig.~\ref{fig:Bc_et}. Users may employ AI technologies to engage in the metaverse interactions and illegally exploit resources, for example, by utilizing AI code to win games or by stealing resources from other users \cite{wiederhold2022ready}. It is also possible that AI will make errors that consequently cause people to lose faith in the metaverse. Furthermore, it is an obstacle to employ a similar type of blockchain across a variety of AI applications in the metaverse.

\subsubsection{How blockchain can help}

The encryption provided by blockchain technology facilitates the metaverse users with complete control over their data, making it simple to transfer ownership of AI consent to another party. Users can persuade applications and others that specific information about them is accurate without disclosing this information to the applications themselves through the use of zero-knowledge proofs, this provides the right to use data for AI model training. As a common feature, blockchain ledgers provide an audit trail, which can be used to check the accountability of all transactions occurring in the metaverse. A zero-knowledge evidence system enables individuals to identify critical facts in the metaverse while also protecting their privacy and retaining ownership of their resources from deepfakes \cite{hussain2021artificial}. This will prevent AI from exploiting resources in the metaverse.

\subsubsection{Summary}

The combination of AI and blockchain will protect the highly sensitive data that AI-driven systems must acquire, store, and use. The sensitive data and information, from coarse to fine, in the metaverse are substantially better protected as a result of this approach. Public blockchains are secure and have authentic data processing, but the collected data are open to all stakeholders in the metaverse. This could be a source of concern and will also harm the AI models in the metaverse. The attackers will exploit the weakness of AI if there are no blockchain standards or regulations in place. Remarkably, it is necessary to introduce a cross-chain converter that enables AI applications to be familiar with the metaverses built on different blockchains.

\subsection{Blockchain for Big Data in the Metaverse}

\subsubsection{Introduction}

Data in the metaverse is more diverse and arrives in greater volumes and with high velocity than data in the real world.
When new data sources in the metaverse are discovered, big data techniques in the metaverse will be used to assemble larger and more sophisticated datasets. The massive amount of data generated by the metaverse will be useful for a wide range of activities, ranging from customer service to data analytics \cite{han2021news}. Big data will provide new insights that will lead to the creation of new opportunities and business models in the metaverse.

\subsubsection{Challenges related to Big Data  in the Metaverse}

Although data storage technologies have advanced, the amount of data has progressively doubled and will continue to increase in the metaverse. Keeping up with the amount and speed at which data is generated in the metaverse is a complicated task to accomplish. The heterogeneity of the data generated by the metaverse applications is also a big challenge. The ability to put data to good use in the metaverse is what makes it valuable, and curation is the method by which this is accomplished. Data collection and organization that is essential to the consumer in the metaverse requires a large investment of time and effort. Data scientists will spend the majority of their time preparing and organizing data for it to be used by stakeholders \cite{wang2022metasocieties}. Finally, big data technology is progressing at a rapid speed. Keeping up with the latest big data technological developments in the metaverse is a never-ending challenge.

\subsubsection{How blockchain can help}

The use of blockchain technology will aid in the collection of data from trusted data sources, hence reducing the amount of improper data obtained as depicted in Fig.~\ref{fig:Bc_et}. The data owners will have total control over their data, and any data manipulation by a third party will be restricted. This assures that data flows in the metaverse are of a high standard of quality \cite{deepa2022survey}. Due to the decentralized nature of blockchain technology, data scientists in the metaverse will be able to communicate and collaborate on data cleaning, which will significantly reduce the time and expense associated with classifying data and creating datasets for analytics applications, as well as the risk of data contamination. Because of the immutability of the blockchain, it will not be feasible to tamper with the data because it will be duplicated throughout the network \cite{gligor2021theorizing}. This will improve the availability of data for stakeholders of the metaverse.

\subsubsection{Summary}

Blockchain holds a great deal of potential for the future of big data analytics. Users will be able to maintain complete control over their personal information and financial activities in the metaverse. There will be no need for a third party to obtain trusted data and to label that data because of the blockchain. Some of the issues, like consensus models, the cost of mining blocks, and the verification of transactions are still challenging \cite{deepa2022survey}. Blockchain  offers solutions that require major changes to existing systems or the complete replacement of these systems. As a result, it will be hard and time-consuming to change the whole system. Even though blockchain integration with the metaverse is still in its early stages, these issues will be resolved in the future, paving the way for a wide range of new and exciting opportunities.

\subsection{Blockchain Multi-sensory XR Applications, Holographic Telepresence in the Metaverse}

\subsubsection{Introduction}

The metaverse provides immersive, and real-world experiences through the use of technology like holographic telepresence and augmented reality applications as depicted in Fig.~\ref{fig:Bc_et}. These applications incorporate audio, video, cognition, and other components. They provide a real-time representation of virtual and physical objects in the metaverse. XR applications will make use of sensors to create a more realistic experience by incorporating real-world objects \cite{yang2021utilization}. As a result of these advancements, holographic telepresence and multi-sensory XR applications enable a user to experience both the real and virtual worlds concurrently.

\subsubsection{Challenges related to Multi-sensory XR Applications, Holographic Telepresence in the Metaverse}

XR technologies like VR, AR, and holographic telepresence are key enabling technologies in the metaverse. However, they may also raise personal and societal concerns. Using the information gathered from these technologies, the companies will be able to develop a recommendation system. The quality of these recommendation systems in the metaverse can be influenced by behavioral data collected from a variety of sources. This technology necessitates enormous amounts of data storage, which must be readily accessible at all times for the users in the metaverse. Sensitive data like biometric information collected by AR/VR devices can be used to identify users and infer additional information about them in the virtual world \cite{duan2021metaverse}. The metaverse must ensure the privacy of such sensitive information of the users.
These gadgets exchange or transfer a massive amount of data between virtual worlds. The metaverse must enable data transparency when various stakeholders and third parties are involved in the data sharing process \cite{bhattacharya2021coalition}. 
\subsubsection{How blockchain can help}

In the metaverse, a blockchain-based distributed ledger would enable the validation of holographic telepresence and other XR applications records, as well as trace the source of erroneous data. This will help build a more accurate recommendation system. Holographic telepresence and other XR applications will find it easier to share data securely between virtual worlds with the zero-trust mechanism and cross-chain technology of the blockchain \cite{bhattacharya2021coalition}. For XR applications and holographic telepresence, the interplanetary file system provided by the blockchain ensures data integrity. The data collected by these devices and saved on a blockchain will be immutable due to the consensus mechanism. Blockchain ensures trust among AR/VR stakeholders by making the verification and ownership transfer of digital assets transparent \cite{kumar2021decentralized}.

\subsubsection{Summary}

Multi-sensory XR applications combined with holographic telepresence and blockchain technology will help to integrate digital economies into unified platforms where assets and payments in the metaverse can be managed efficiently and unambiguously. VR/AR technology will not have the global reach of smartphones or computers in the foreseeable future. A concern and a problem for blockchain will be the use of enhanced AI deep fakes \cite{johnson2021deepfakes}, which must be addressed by new emerging blockchain platforms.

IoT devices, digital twins, and XR applications will generate big data in the metaverse. Blockchain will assist these technologies in high quality, secured acquisition of authentic data in the metaverse. Blockchain will store and handle the big data in the metaverse data in a secured manner through the immutability and transparency properties offered by the blockchain. Enabling technologies like IoT, digital twins, and XR will be benefited considerably in terms of higher data transparency and adaptability provided by the blockchain. NFTs and the real estate or the digital assets produced using digital twins in the metaverse will become interoperable because of the zero-trust and cross-chain mechanisms of the blockchain. The metaverse-enabled technologies will be benefited from the high degrees of data privacy due to anonymity, audits, authentication, anti-tampering, and access control offered by blockchain. Table \ref{tab:Blockchain} summarizes the impact of the blockchain on the technical aspects and various enabling technologies of the metaverse.

\begin{table}[!t]
\centering
    \caption{Impact of Blockchain for Technical Aspects in the Metaverse }
    \label{tab:Blockchain}
\renewcommand{\arraystretch}{1.2}     
\begin{tabular}
  {|p{3cm}|c|c|c|c|c|c|}  
  \rowcolor{gray!25}
 
  \hline
   \cellcolor{gray!15} Technical Perspective &{\rotatebox[origin=c]{90}{\cellcolor{gray!15}~Data Acquisition Capacity~}}
   &{\rotatebox[origin=c]{90}{\cellcolor{gray!15}~Data Storage~}}
   &{\rotatebox[origin=c]{90}{\cellcolor{gray!15}~Data Sharing~}}
   &{\rotatebox[origin=c]{90}{\cellcolor{gray!15}~Data Interoperability~}}
   &{\rotatebox[origin=c]{90}{\cellcolor{gray!15}~Data Privacy Preservation~}}
    \\
\hline   
\hline
   IoT  & \cellcolor{green!15}H & \cellcolor{red!15}L & \cellcolor{green!15}H & \cellcolor{red!15}L & \cellcolor{green!15}H  \\
\hline
  Digital Twins & \cellcolor{green!15}H & \cellcolor{red!15}L & \cellcolor{green!15}H & \cellcolor{yellow!15}M & \cellcolor{green!15}H  \\
 \hline
   XR Applications & \cellcolor{green!15}H & \cellcolor{yellow!15}M & \cellcolor{green!15}H & \cellcolor{red!15}L & \cellcolor{green!15}H\\
\hline
   AI & \cellcolor{red!15}L & \cellcolor{yellow!15}M & \cellcolor{red!15}L & \cellcolor{red!15}L & \cellcolor{green!15}H \\
\hline
   Big Data  & \cellcolor{green!15}H & \cellcolor{green!15}H & \cellcolor{red!15}L & \cellcolor{red!15}L & \cellcolor{green!15}H \\
\hline
 
\end{tabular}
\begin{flushleft}
\begin{center}
    
\begin{tikzpicture}

\node (rect) at (0,2) [draw,thick,minimum width=0.6cm,minimum height=0.4cm, fill= red!15, label=0:Low Impact] {L};
\node (rect) at (2.8,2) [draw,thick,minimum width=0.6cm,minimum height=0.4cm, fill= yellow!15, label=0:Medium Impact] {M};
\node (rect) at (6,2) [draw,thick,minimum width=0.6cm,minimum height=0.4cm, fill= green!15, label=0:High Impact] {H};
\end{tikzpicture}
\end{center}

\end{flushleft}
  
\end{table}





\section{The Metaverse Projects}
\label{sec:Project}

\begin{figure*}[!t]
	\centering
	\includegraphics[width=180mm]{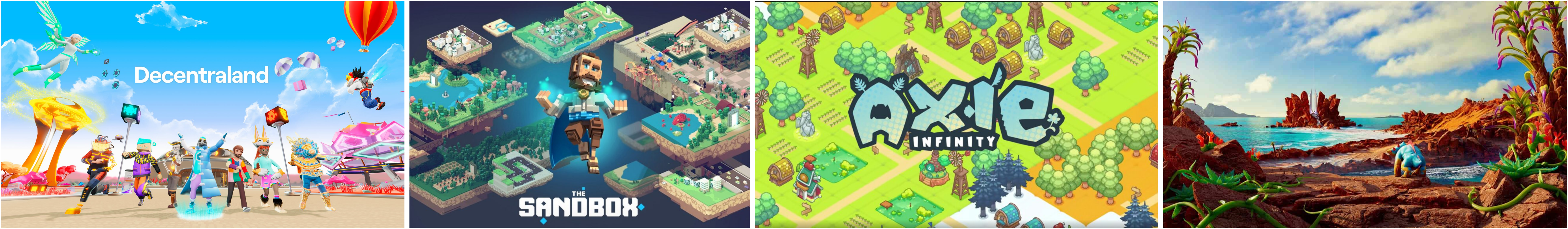}
	\caption{Inside the virtual worlds of different metaverse projects (left to right): Decentraland, Sandbox, Axie Infinity, and Illuvium.}
	\label{fig_project}
\end{figure*}

This section briefly introduces some well-known the metaverse projects: Decentraland, Sandbox, Axie Infinity, and Illuvium, which have exploited blockchain as the core technology of the metaverse foundation and development, and additionally to deliver multifarious blockchain-based services and applications in the virtual world, from real estate to E-commerce and real estate. The landscapes inside the virtual worlds of the projects are shown in Fig.~\ref{fig_project}.

\textit{Decentraland\footnote{https://decentraland.org/}}: 
Decentraland is a virtual reality platform powered by the Ethereum blockchain, which allows users to experience, create, and monetize economic assets, hyperreal contents, and applications. Land in Decentraland is permanently owned by the community with full surveillance and control over their creative activities. In the virtual world of Decentraland, a virtual land is identified uniquely as a non-fungible, transferrable, and scare digital asset stored in the Ethereum smart contract, which can be claimed on the blockchain-based ledger with ERC20 (Ethereum Request for Comments 20) tokens call MANA.
Different from other traditional virtual worlds and social networks, Decentraland is not managed and supervised by any centralized organization; that is, no single agent has permission to change the rules of software, content, economic mechanism, or prevent others from accessing the world, trading digital products, and providing services. 

Decentraland is with the proof of concept to enable the ownership of digital real estate to the user on the Ethereum blockchain, wherein its blockchain payment network is built to obtain short-to-medium-term scalability. This payment network is with low fees to encourage the economic development of the Decentraland metaverse. In the perspective of user cases, Decentraland considers content curation, advertising, social (forums, chat groups, and multiplayer games), applications (games, gambling, and dynamic 3D scenes with scripting language toolset), and others (such as therapy, 3D design, education, and virtual tourism). In the perspective of architecture, the Decentraland protocol is comprised of three layers: consensus layer, land content layer, and real-time layer besides two supportive systems, including payment channel infrastructure and identity system. With the Ethereum smart contract to maintain a ledger of ownership for a piece of land in the virtual world, Decentraland remarks non-fungible digital assets LAND which is generated by burning MANA tokens via the LAND smart contract. For ownership identification, Decentraland uses the Ethereum Name Service to create of a layer of ownership over in-world items.

\textit{Sandbox\footnote{https://www.sandbox.game/en/}}: 
As built on the Ethereum blockchain, Sandbox is a user-generated the decentralized metaverse, in which users can build, own, and monetize immersed gaming experiences using SAND, its native platform’s utility token with ERC-20. 
Inspired by Minecraft\footnote{https://www.minecraft.net/en-us} and Roblox\footnote{ https://www.roblox.com/}, Sandbox uplifts the gaming experience from a 2D mobile pixel environment to a fully 3D world by using a voxel gaming platform. In the Sandbox metaverse, users can freely create and animate 3D objects (such as people, animals, buildings, and tools) by VoxEdit, a built-in voxel gaming package, with true ownership as NFTs. These creations can be traded on the Sandbox marketplace as game assets and the creators can receive their reward/incentive by SAND tokens (compatible with ERC-721 and ERC-1155).
With the utility of NFTs, the Sandbox users will receive the following benefits: (i) true digital ownership (every game item can be tokenized to easily own, trade, and sell), (ii) security and immutability (the risks of fraud and theft are minimized with the distributed ledger of blockchain technology), (iii) trading (ultimate digital asset control without in-game value abandon), and (iv) cross-platform interoperability (in-game items can be available for usage in different games that allow it).
The Sandbox metaverse adopts The InterPlanetary File System to store the actual digital assets and ensure the assets can be modified without owner permission. The main concern of Ethereum chain-aided projects is scalability, which motivates the Sandbox team to look at layer-2 solutions (scale an application by handling transactions off the Ethereum Mainnet while taking advantage of decentralized security). 

\textit{Axie Infinity\footnote{https://axieinfinity.com/}}: 
As one of the revolutionary Play-to-Ern metaverse projects, Axie Infinity builds a crypto-meet-Pok\'emon game universe with fantasy creatures, called Axies, that players can collect, raise, breed, and battle for building their Axies kingdoms. Like Decentraland and Sandbox, Axie Infinity has a user-centric economy system that allows players to truly own, sell, buy, and trade in-game resources over gaming activities and contributions to the ecosystem. A key difference between Axie and other traditional games is that the blockchain-based economic mechanism of the Axie metaverse allows players to increase their digital assets by improving in-game skills to reach certain levels. Players can have fun with many play modes (PvP - player versus player and PvE - player versus environment) and numerous tournaments while earning in-game resources for trading with real monetary value possibly. 
Axie Infinity Shards (AXS), the ERC-20 governance token of the Axie metaverse, can be claimed as rewards when players stake their AXS tokens, play the game, and participate in governance activities. Furthermore, they can earn AXS tokens when playing different games involved by Axie Infinity Universe and creating user-generated content. Axies creatures and other virtual real estates can be bought, sold, and traded via an in-game marketplace in the form of NFTs. Remarkably, most transactions are processed on an Ethereum-linked sidechain, called Ronin, which is specially designed to achieve lower fees than the standard Ethereum blockchain. Besides AXS, Axie Infinity has a secondary token, namely Small Love Position (SLP), which is awarded to players via in-game activities.

\textit{Illuvium\footnote{https://www.illuvium.io/}}: 
Often touted as the first open-world fantasy battle game metaverse built on the Ethereum blockchain, Illuvium can provide a source of entertainment to regular gamers and decentralized finance (DeFi) users through a range of collecting and trading features. The virtual world of Illuvium is inhabited by fantasy creatures, called Illuvials, which can be captured by players when defeating them in casual battles. From then on, these Illuvials become a loyal team of the player’s collection and be carried out to combat against other players via a random PvP gameplay. In other words, the Illuvium’s game is a combination of an open-world exploration game and a PvP battle game, where players can immerse different gameplays, i.e., freely exploring the virtual world while planning battle statics. Each Illuvial is associated with a unique NFT and can be traded on the in-game marketplace or external exchange platform without gas fees. 

The native token used within the Illuvium ecosystem is ILV which has three main use cases: rewarding players for in-game achievements, presenting players for their vault (a private wallet of each user) distribution, and participating in governance activities of the game via decentralized autonomous organization (DAO). To obtain the scalability of applications with NFT functionality, Illuvium leverages Immutable X, a layer-2 Ethereum scaling solution that allows users to trade NFTs with zero gas fees and instant transaction finality by using an innovative technique known as Zero-Knowledge Rollup. Besides layer-2 integration, Illuvium is featured by a built-in decentralized exchange platform that facilitates trading activities of Illuvial NFTs.

\section{Conclusion and Research Directions}
\label{Sec:Conclusion}
The paper has comprehensively investigated and analyzed the roles and impacts of blockchain for the foundation and development of applications and services in the metaverse. 
The fundamental concepts of blockchain and the metaverse were sketched at the beginning of this work, along with the role of blockchain regarding the foundation and development of the metaverse.
Later in this work, several prominent technical aspects and use cases of blockchain in the metaverse were investigated exhaustively besides the insightful challenge analysis and applicability discussion given.
Finally, some technical improvements of blockchain were provisioned for the metaverse, which in turn enhances the performance and practicality of potential applications and services in the virtual world.
Besides making the conclusion, we sketch out some future research directions as below.

Relying on the systematic investigation of blockchain for the metaverse in both the technical and use case perspectives, blockchain had showed a great potential to revolutionize the immersive experience with various applications and services built in the virtual world. Many technical and applicable aspects of all current blockchain versions have been attracting much more research activities, including consensus algorithms, network management, and blockchain interoperability. As consensus algorithms ensure the agreement of states of certain data among authorized nodes in a distributed network, numerous variations of consensus mechanisms have been introduced to achieve high throughput and low latency, but security, scalability, and decentralization could not be obtained concurrently~\cite{bhutta2021survey}. In this context, it is necessary to develop and sharpen some hybrid innovative consensus algorithms (e.g., Proof-of-Capability, Proof-of-Burn, and Leased Proof-of-Stake) to effectively handle the above-mentioned issues. As a serious global issue noticed by many governments and blockchain communities, high energy consumption and greenhouse gas emission derived by the operation of a large number of participating nodes in a network has caused negative climate and environmental impacts. For a sustainable solution, the Stellar consensus protocol~\cite{stellar2015} allows authenticating transactions based on a set of trustworthy nodes rather than running the authentication process for the whole network as PoW or PoS algorithm, which in turn accelerates the speed and reduces energy in use. 

Nowadays, numerous networks and blockchains have been designed for specific applications and services under the umbrella of different community organizations and government departments; therefore, interconnecting existing and new chains is necessary to boost the development of emerging technologies in the metaverse. Cross-chain is introduced as the ultimate solution to obtain the interoperability between different chains, which allows users to execute transactions (with value and information) successfully between different blockchain networks~\cite{geroni2021blockchain}. For instance, users would be able to send the information from an Ethereum blockchain to a Polygon blockchain and vice versa. This interoperability technique also actuates the increasing development of fully decentralized systems with cross-chain bridges. For a long-term evolution of blockchain to reach multi-chain interoperability, omni-chain~\cite{zarick2021layerzero}, which is defined as a blockchain-as-a-service platform to interact with a wide variety of enterprise networks, can provide blockchain-based applications and services (including asset management, smart contract, transaction management, and shared data ledger) with many appreciated benefits, such as greater transparency, enhanced security, improved traceability, and better efficiency and speed.

In traditional organizations adopting classical company hierarchy, most of the important decisions have been made by directors and area managers, which are usually risky and faulty due to human mistakes.
To overcome these problems, DAOs are introduced as the next generation of organizational structure, which involves a group of incognito people sharing the information together according to a self-enforcing protocol~\cite{voshmgir2020dao}.
DAOs are governed by the smart contract algorithms that live on a blockchain network to reduce transaction management fees while presenting better transparency and incorruptibility. 
All governance rules of a DAO are recorded on a transparent, secure, and open-source ledger of the blockchain network. Noticeably, native token stakeholders do not have permission to override rules, but they can use tokens to vote proposals via DAO's consensus rules. 
Concerning shareholders' common goals, the immutability of smart contracts in DAOs will preserve the economic profit and other interests of any governed organizations over a tamper-proof shared ledger, where all activities and transactions on the network will be recorded.
In the future, services (e.g., DeFi) and products (e.g., NFT) in the metaverse can be developed and maintained automatically by DAO, in which smart contracts and consensus rules will govern all major functionalities.




\bibliographystyle{IEEEtran}
\bibliography{reference}

\end{document}